Title: Holography and drag force in thermal plasma of non-commutative 
Yang-Mills theories in diverse dimensions
Authors: Shibaji Roy
Comments: 13 pages

We use holography and a string probe approach to compute the drag force
on a quark moving in a thermal plasma of non-commutative Yang-Mills (NCYM)
theories in various dimensions. The gravity background in these cases are
described by a particular decoupling limit of non-extremal (D$(p-2)$, D$p$)
brane bound state system. We show how the drag force on an external quark
moving in the dual NCYM theories gets corrected due to non-commutativity and
as a result the effective viscosity of the plasma gets reduced. We have
obtained the drag force for both small and large non-commutativity. This was
known earlier for (3+1)-dimensional NCYM theory, however, we find that
the corrections for the general case typically depend on the dimensionality
of the NCYM theories, indicating that the structure of the drag force is
non-universal.

\documentclass[12pt]{article}

\usepackage{amsfonts}
\usepackage[centertags]{amsmath}
\usepackage{amssymb}
\usepackage{cite}
\usepackage{psfrag}
\usepackage{graphicx}
\usepackage{bbm,bm}
\usepackage{epsfig, multicol}

\allowdisplaybreaks
\begin{document}

\topmargin 0pt
\oddsidemargin 0mm
\def\be{\begin{equation}}
\def\ee{\end{equation}}
\def\bea{\begin{eqnarray}}
\def\eea{\end{eqnarray}}
\def\ba{\begin{array}}
\def\ea{\end{array}}
\def\ben{\begin{enumerate}}
\def\een{\end{enumerate}}
\def\nab{\bigtriangledown}
\def\tpi{\tilde\Phi}
\def\nnu{\nonumber}
\newcommand{\eqn}[1]{(\ref{#1})}

\newcommand{\vs}[1]{\vspace{#1 mm}}
\newcommand{\dsl}{\pa \kern-0.5em /} 
\def\a{\alpha}
\def\b{\beta}
\def\g{\gamma}\def\G{\Gamma}
\def\d{\delta}\def\D{\Delta}
\def\ep{\epsilon}
\def\et{\eta}
\def\z{\zeta}
\def\t{\theta}\def\T{\Theta}
\def\l{\lambda}\def\L{\Lambda}
\def\m{\mu}
\def\f{\phi}\def\F{\Phi}
\def\n{\nu}
\def\p{\psi}\def\P{\Psi}
\def\r{\rho}
\def\s{\sigma}\def\S{\Sigma}
\def\ta{\tau}
\def\x{\chi}
\def\o{\omega}\def\O{\Omega}
\def\k{\kappa}
\def\pa {\partial}
\def\ov{\over}
\def\nn{\nonumber\\}
\def\ud{\underline}
\begin{flushright}
arXiv:YYMM.NNNN\\
\end{flushright}
\begin{center}
{\large{\bf Holography and drag force in thermal plasma of 
non-commutative Yang-Mills theories in diverse dimensions}}

\vs{10}

{Shibaji Roy\footnote{E-mail: shibaji.roy@saha.ac.in}}

\vs{4}

{Saha Institute of Nuclear Physics\\
1/AF Bidhannagar, Calcutta 700064, India\\}

\end{center}

\vs{15}

\begin{abstract}

We use holography and a string probe approach to compute the drag force
on a quark moving in a thermal plasma of non-commutative Yang-Mills (NCYM)
theories in various dimensions. The gravity background in these cases are
described by a particular decoupling limit of non-extremal (D$(p-2)$, D$p$)
brane bound state system. We show how the drag force on an external quark
moving in the dual NCYM theories gets corrected due to non-commutativity and
as a result the effective viscosity of the plasma gets reduced. We have 
obtained the drag force for both small and large non-commutativity. This was 
known earlier for (3+1)-dimensional NCYM theory, however, we find that 
the corrections for the general case typically depend on the dimensionality 
of the NCYM theories, indicating that the structure of the drag force is 
non-universal.  

\end{abstract}

\newpage

The AdS/CFT correspondence \cite{Maldacena:1997re,Witten:1998qj,Gubser:1998bc} 
or its generalized versions (for a review see \cite{Aharony:1999ti}) 
holographically 
relate string theory in a particular background to a specific gauge theory. 
If the
string theory is weakly coupled then the corresponding gauge theory is strongly
coupled ('t Hooft coupling $\lambda = g_{YM}^2 N \gg 1$) and {\it
vice-versa}. This 
gives an access or computational handle on the 
strongly coupled gauge theory from weakly coupled string theory or 
supergravity 
when the higher curvature corrections are also under control. This is precisely
the idea behind several interesting predictions for some observables which
might be related to quark gluon plasma (QGP), for example, 
the entropy production, transport
properties, jet quenching etc. (for a recent review see 
\cite{Shuryak:2008eq}).  
Collisions of heavy nuclei in the lab (such as
at RHIC or in the near future at LHC) are believed to produce 
QGP, a thermal state of matter, which 
is strongly coupled and behaves like an ideal fluid \cite{Shuryak:2003xe}. 
This, therefore, provides
a good laboratory to test the (generalized) AdS/CFT correspondence and indeed
many such calculations on the QGP observables just mentioned have been
performed using AdS/CFT correspondence and compared with the experimental 
results with some partial success 
(see \cite{Liu:2006he,Shuryak:2008eq,Gubser:2009md} and references therein). 
In the original calculation \cite{Policastro:2001yc} a non-extremal
D3-brane solution of type IIB string theory was used to compute the shear
viscosity of the thermal YM gauge theory on the boundary using AdS/CFT
correspondence. The non-extremal solutions of string theory are believed to
be dual to thermal gauge theories with properties similar to QCD at strong
coupling. 

Subsequently, in this spirit, many such computations
have been performed in
various strongly coupled thermal gauge theories (with 
\cite{Policastro:2002se,Policastro:2002tn,Kovtun:2004de,Gubser:2006bz,Liu:
2006ug} or without conformal
symmetries \cite{Benincasa:2005iv,Buchel:2006bv,Mas:2007ng}, 
chemical potentials \cite{Son:2006em,Mas:2006dy,Benincasa:2006fu}, 
finite 't Hooft coupling \cite{Armesto:2006zv,Buchel:2008wy,Buchel:2008ac} 
etc.), whose gravity
duals are given by a set of coincident non-extremal D3-branes, D$p$-branes,
rotating D3-branes etc. They were even extended to other geometries having
duals like Klebanov-Witten CFT, Leigh-Strassler, Klebanov-Strassler cascading 
gauge theories and others \cite{Liu:2006he,Buchel:2005cv,Buchel:2009bh,Buchel:2004hw}.
One of the interesting observations in these calculations is the universality
of the ratio of shear viscosity to entropy density ($\eta/s$) and its bound
\cite{Kovtun:2004de,Buchel:2003tz} (see \cite{Son:2007vk} for a recent status
on this). 
However, no such universality has been observed in the calculation of jet 
quenching parameter ($\hat{q}$), a measure of the radiative parton energy
loss in a medium \cite{Liu:2006ug,Liu:2006he}. For example, for the 
non-conformal, thermal gauge theories 
living on the boundary of a stack of coincident non-extremal D$p$-branes
\cite{Itzhaki:1998dd}, 
the signature of the non-universality can be seen to be encoded in the 
expression of jet
quenching parameter as $\hat{q} \sim T^2 (T \sqrt{\lambda})^{\frac{2}{5-p}}$,
where $T$ is the temperature of the gauge theory and $\lambda$ is the 't Hooft
coupling \cite{Liu:2006he}. So, both the powers of $T$ and $N$ are 
dependent on the
dimensionality of the gauge theory and this is the reason there is no
universality in its structure. Another interesting quantity is the viscous
drag force experienced by an external quark moving in a hot plasma 
\cite{Gubser:2006bz,Herzog:2006gh,CasalderreySolana:2006rq}. 
In the
D-brane picture an external quark is represented as the end point of a
fundamental string attached to the boundary carrying a fundamental charge
under the gauge group SU($N$) and is infinitely massive 
\cite{Maldacena:1998im,Rey:1998bq,Rey:1998ik,Brandhuber:1998bs}. 
The external quark
loses its energy as the string attached to it trails back and imparts a drag
force on it. Though originally \cite{Gubser:2006bz} it was calculated from the 
motivation to understand the phenomenon
of jet quenching in the medium produced in heavy ion collision, it was
later realized \cite{Liu:2006he} 
that they are not quite related to each other. The drag force,
also like the jet quenching parameter, is non-universal and for D$p$-branes
the force is given as $F \sim T (T \sqrt{\lambda})^{\frac{2}{5-p}}$ 
\cite{Liu:2006he}. 

In this short note we study the effect on the drag force experienced by a
quark when it moves through a hot plasma having a space-space
non-commutativity. It is well-known that the gauge theory develops a
space-space non-commutativity when D$p$-brane world-volume is subjected to
a large asymptotic magnetic or $B$-field 
\cite{Seiberg:1999vs,Maldacena:1999mh,Hashimoto:1999ut}. 
This effect has been studied earlier
in \cite{Matsuo:2006ws} for D3-branes and it was found that the drag force 
on top of the
value $\sim \sqrt{\hat\lambda} T^2$ gets corrected by 
$ \sim -\hat\lambda^{3/2} T^6
\theta^2$, for small $\theta$, where $\theta$ is the non-commutativity 
parameter and $\hat\lambda$ is the 't Hooft coupling for the NCYM theory. 
Whereas, for large non-commutativity parameter, $\theta \gg 1$, 
the drag force is given as $\sim (\sqrt{\hat\lambda} T^2 \theta^2)^{-1}$ in the
leading order and $\sim -(\hat\lambda^{3/2} T^6 \theta^4)^{-1}$ in the next to
leading order. Note that for large non-commutativity the dependence on
$\hat\lambda$ and $T$ got inverted in the expression of drag 
force from those for 
small non-commutativity. Also note that the drag force gets reduced by the
effect of non-commutativity and so, the noncommutativity reduces the viscous
force. Now in order to see, whether there is any universal structure we
compute in this note the expression of the drag force on an external quark 
for the hot $(p+1)$ dimensional YM plasma having space-space
non-commutativity. We obtain the drag force for both small and large 
non-commutativity and, like jet quenching parameter and the drag force without
non-commutativity, we find that the expression typically depends on the
dimensionality of space-time and so does not possess a universal structure.
The shear viscosity in the non-commutative plasma has been discussed in
\cite{Landsteiner:2007bd}.

The finite temperature $(p+1)$-dimensional NCYM theory is holographically
dual to the stack of non-extremal (D$(p-2)$, D$p$) supergravity bound state
solutions of type II string theory in a particular decoupling limit 
\cite{Maldacena:1999mh,Alishahiha:1999ci}. 
The
world-volume of D$(p-2)$ lies completely within the worldvolume of D$p$-branes
and their extremal version has 16 supercharges. The complete non-extremal
(D$(p-2)$, D$p$) bound state solution is given as \cite{Cai:2000hn},
\bea\label{dpdp2}
ds^2 &=& H^{-\frac{1}{2}}\left[-f dt^2 + \sum_{i=1}^{p-2} (dx^i)^2 + 
\frac{H}{F} \left((dx^{p-1})^2 + (dx^p)^2\right)\right] +
H^{\frac{1}{2}}\left[\frac{dr^2}{f} + r^2 d\Omega_{8-p}^2\right]\nn
e^{2(\phi - \phi_0)} &=& \frac{H^{\frac{5-p}{2}}}{F}, \qquad 
B_{p-1,\,p} = \frac{\tan\alpha}{F}\nn
A_{012\ldots p} &=& \frac{1}{g_s} \frac{\left(1 -H \right)}{F}
\cos\alpha \coth\varphi, \qquad  A_{012\ldots p-2} = \frac{1}{g_s} 
\left(H^{-1} -1\right)\sin\alpha \coth\varphi
\eea    
where the various functions appearing above are defined as,
\bea\label{functions}
f &=& 1 - \frac{r_0^{7-p}}{r^{7-p}}\nn
H &=& 1 + \frac{r_0^{7-p} \sinh^2\varphi}{r^{7-p}}\nn
F &=& 1 + \frac{r_0^{7-p} \sinh^2\varphi \cos^2\alpha}{r^{7-p}}
\eea
Here the D$p$-branes lie along $x^1,\ldots,x^p$ and the D$(p-2)$
branes lie along $x^1,\ldots,x^{p-2}$. The angle $\alpha$ measures the
relative numbers of the D$(p-2)$ and D$p$ branes and is defined as,
$\cos\alpha = N/\sqrt{N^2+M^2}$, where $N$ is the number of D$p$-branes and 
$M$ is number of D$(p-2)$-branes per unit co-dimension two volume transverse to
D$(p-2)$ brane. Also $\varphi$ is the boost parameter and $r_0$ is the radius
of the horizon of the non-extremal or black (D$(p-2)$, D$p$) solution. The NCYM
decoupling limit is a low energy limit by which we zoom into the region given
by \cite{Maldacena:1999mh,Alishahiha:1999ci}
\be\label{NCYMlimit}
r_0 < r \sim r_0 \sinh^{\frac{2}{7-p}} \varphi \cos^{\frac{2}{7-p}}
\alpha \ll r_0 \sinh^{\frac{2}{7-p}}\varphi 
\ee
It is clear from above that $\varphi$ is very large whereas $\alpha$
is an angle very close to $\pi/2$. In this approximation 
\bea\label{approxfns}
H & \approx & \frac{r_0^{7-p} \sinh^2\varphi}{r^{7-p}}\nn
\frac{H}{F} &\approx & \frac{1}{\cos^2\alpha (1+a^{7-p} r^{7-p})}
\equiv \frac{h}{\cos^2\alpha} 
\eea
where,
\be\label{definitions} 
h = \frac{1}{1+a^{7-p}r^{7-p}}, 
\qquad {\rm with} \qquad a^{7-p} = \frac{1}{r_0^{7-p} \sinh^2\varphi
\cos^2\alpha}    
\ee 
We see from \eqn{dpdp2} that the asymptotic value of the $B$-field is
$\tan\alpha$ and since $\alpha$ is close to $\pi/2$, it is very large in 
the NCYM limit. In this limit the metric in \eqn{dpdp2} takes the form,
\be\label{newmetric}
ds^2 = H^{-\frac{1}{2}}\left[-dt^2 + \sum_{i=1}^{p-2}(dx^i)^2 +
h\left((dx^{p-1})^2 + (dx^p)^2\right)\right] + H^{\frac{1}{2}}\left[
\frac{dr^2}{f} + r^2 d\Omega_{8-p}^2\right]
\ee
where $H$ and $h$ are as given in \eqn{approxfns} and we have rescaled the
coordinates $x^p$ and $x^{p-1}$ as $x^{p-1,p} \to \cos\alpha\,\, x^{p-1,p}$. 
In order to calculate
the drag force on an external quark moving through the hot NCYM plasma, we
look at the dynamics of the fundamental string in the background 
\eqn{newmetric} given by the Nambu-Goto action,
\be\label{action}
S = -\frac{1}{2\pi\alpha'}\int d\tau d\sigma \sqrt{-{\rm det} (g_{ab})}
\ee
where $g_{ab}$ is the induced metric given by,
\be\label{inducedmetric}
g_{ab} = \frac{\partial X^\mu}{\partial \xi^a}\frac{\partial X^\nu}
{\partial \xi^b} G_{\mu\nu}
\ee
$G_{\mu\nu}$, in the above is the background metric \eqn{newmetric} and 
$\xi^{a,b}, a,b = 0,1$, are the world-sheet coordinates $\tau = \xi^0$ and
$\sigma = \xi^1$. We use the static gauge condition $X^0 \equiv t = \tau$ and 
$r = \sigma$ and the string is allowed to move along one of the
non-commutative directions $X^p = x$. Then the string embedding is completely
specified by the function $x(t,r)$. The action \eqn{action} then reduces to
the form,
\be\label{action1}
S = -\frac{1}{2\pi\alpha'} \int dt dr \left[1 - \frac{h}{f}(\dot x)^2 +
\frac{f h}{H} (x')^2\right]^{\frac{1}{2}}
\ee     
where the `overdot' represents derivative with respect to `$t$' and `prime' 
denotes
derivative with respect to `$r$'. Since the action does not depend
explicitly on `$t$' and `$r$', we get two constants of motion,
\bea\label{constofmotion1}
\pi_x^0 &=& \frac{1}{2\pi\alpha'}\frac{h \dot x}{f \sqrt{1 - 
\frac{h}{f}(\dot x)^2 +
\frac{f h}{H} (x')^2}} = {\rm const.\,\,indep.\,\,
of\,\,}t\\\label{constofmotion2}
\pi_x^1 &=& -\frac{1}{2\pi\alpha'}\frac{f h x'}{H \sqrt{1 - 
\frac{h}{f}(\dot x)^2 +
\frac{f h}{H} (x')^2}} = {\rm const.\,\,indep.\,\,
of\,\,}r
\eea
Now we make a simplifying assumption that if we allow sufficiently long time
(since the quark is heavy) the quark or the string will eventually move with 
a constant velocity (quark will interact with the plasma and lose its energy)
$v$, i.e., 
\be\label{xprofile}
x(t,r) = v t + \zeta(r)
\ee
Substituting \eqn{xprofile} into \eqn{constofmotion1} and
\eqn{constofmotion2}, we find that \eqn{constofmotion1} is automatically 
satisfied and \eqn{constofmotion2} gives,
\be\label{const}
\pi_{\zeta} = \frac{f h \zeta'}{H \sqrt{1 - \frac{h}{f} v^2 + 
\frac{f h}{H}(\zeta')^2}} = {\rm const.\,\, indep.\,\, of\,\,} r,t
\ee
Solving \eqn{const} we obtain,
\be\label{solution}
\zeta' = \frac{H \pi_{\zeta}}{h f} \sqrt{\frac{\left(\frac{f}{h} - v^2\right)}
{\left(\frac{f}{h} - \frac{H \pi_{\zeta}^2}{h^2}\right)}}
\ee
In \eqn{solution} we have assumed $\zeta'$ to be positive. Note that as $r$
varies from $\infty$ to $r_0$, both the numerator and the denominator in the
square-root in \eqn{solution} change sign and therefore $\zeta'$ can become 
imaginary for some
values of $r$. So, this solution is not physically acceptable. The
easiest way to avoid this problem is to choose the constant $\pi_{\zeta}$,
such that both the numerator and the denominator change sign at the
same place $r_v$ (say). This requirement fixes the value of $\pi_{\zeta}$ 
to be of the form,
\be\label{pi}
\pi_{\zeta} = \frac{v r_v^{\frac{7-p}{2}}}{
(1+a^{7-p} r_v^{7-p}) r_0^{\frac{7-p}{2}} \sinh\varphi}
\ee
where 
\be\label{rv}
r_v^{7-p} = \frac{-\left(1-v^2-a^{7-p} r_0^{7-p}\right) + \sqrt{\left(1-v^2-
a^{7-p}r_0^{7-p}\right)^2 + 4 a^{7-p}r_0^{7-p}}}{2 a^{7-p}}
\ee
Now we can substitute the value of $\pi_{\zeta}$ from \eqn{pi} into 
\eqn{solution} and integrate to find out the string profile $x(t,r)$. In 
principle this can be done, but actually, in this case we can not write 
$x(t,r)$ in a closed form. On the other hand, the drag force can be easily
calculated as,
\be\label{dragforce}
F = - \pi_x^1 = \frac{1}{2\pi\alpha'}  \frac{v r_v^{\frac{7-p}{2}}}{
(1+a^{7-p} r_v^{7-p}) r_0^{\frac{7-p}{2}} \sinh\varphi}
\ee
where $r_v$ is as given above in \eqn{rv}. It is not not easy to see 
how non-commutativity affects the drag force from 
the general expression given in \eqn{dragforce}. So, for this purpose,
we will consider two limiting cases from which the effect of
non-commutativity on the drag force will be transparent. Therefore, we consider
(i) $ar_0 \ll 1$ (as we will see this corresponds to small non-commutativity)
and (ii) $ar_0 \gg 1$ (this case corresponds to large non-commutativity).

(i) for $ar_0 \ll 1$, the expression of the drag force \eqn{dragforce} can
be simplified to give,
\be\label{dragforcesmall}
F = \frac{1}{2\pi\alpha'} \frac{v}{\sqrt{1-v^2}} \frac{1}{\sinh\varphi}\left[
1 - \frac{2-v^2}{2(1-v^2)^2} a^{7-p}r_0^{7-p} + O(a^{2(7-p)}r_0^{2(7-p)})
\right]
\ee

and (ii) for $ar_0 \gg 1$, \eqn{dragforce} can be simplified as,
\be\label{dragforcebig}
F = \frac{1}{2\pi\alpha'} \frac{1}{\sinh\varphi} \frac{v}{a^{7-p}r_0^{7-p}}
\left[1 - \frac{2+v^2}{2 a^{7-p}r_0^{7-p}} + 
O(\frac{1}{a^{2(7-p)}r_0^{2(7-p)}})\right]
\ee
In eqs.\eqn{dragforcesmall}, \eqn{dragforcebig}, we obtained 
the expression
of the drag force in terms of the parameters of the gravity theory. In order
to understand the nature of the force in terms of the NCYM theory we have to
relate the parameters of the gravity theory to those of the NCYM theory. We
first mention that the non-extremal (D$(p-2)$, D$p$) solution in the NCYM 
limit given in \eqn{newmetric} has a Hawking temperature which by holography
is the same as the temperature of the NCYM theory. The temperature can be
calculated from the metric in \eqn{newmetric} as,
\be\label{temperature}
T = \frac{7-p}{4\pi r_0 \sinh\varphi}
\ee
Also from the charge of the D$p$-brane we can calculate
\be\label{2ndrelation}
r_0^{7-p} \sinh^2\varphi = d_p \hat{\lambda} \alpha'^{5-p}
\ee
where $d_p = 2^{7-2p} \pi^{\frac{9-3p}{2}}\Gamma((7-p)/2)$ is a $p$-dependent
constant and $\hat{\lambda} = \hat{g}_{YM}^2 N$, is the 't Hooft coupling
for the NCYM theory and $\hat{g}_{YM}$ is the NCYM coupling, with $N$ being
the number of D$p$-branes. We point out that the 't Hooft coupling
$\hat{\lambda}$ for the NCYM theory differs from the ordinary YM theory by a
scaling of the form $\lambda = (\alpha'/\theta) \hat{\lambda}$, where $\theta$
is the non-commutativity parameter given as $[x^{p-1},\,\, x^p]= i\theta$.
Here $\theta$ is a finite parameter and in the decoupling limit as $\alpha'
\to 0$, $\hat{g}$ remains finite \cite{Maldacena:1999mh}.
Now using the two relations \eqn{temperature} and \eqn{2ndrelation} we obtain,
\bea\label{parameter1}
\sinh\varphi &=& \left[\frac{(7-p)^{7-p}}{(4\pi T)^{7-p} d_p \hat{\lambda}
\alpha'^{5-p}}\right]^{\frac{1}{5-p}}\\\label{parameter2}
r_0 &=& \left[\frac{(4\pi T)^2}{(7-p)^2} d_p \hat{\lambda} 
\alpha'^{5-p}\right]^{\frac{1}{5-p}}
\eea  
Also from \eqn{definitions} we have 
\be\label{ar}
a^{7-p} r_0^{7-p} = \frac{1}{\sinh^2\varphi \cos^2\alpha} = 
\frac{(4\pi T)^{\frac{2(7-p)}{5-p}} d_p^{\frac{2}{5-p}} 
\hat{\lambda}^{\frac{2}{5-p}} \theta^2}{(7-p)^{\frac{2(7-p)}{5-p}}}
\ee
In the above we have used that in the decoupling limit $\cos\alpha =
\alpha'/\theta$, i.e. as $\a' \to 0$, $\alpha \to \pi/2$. Also note from 
\eqn{ar} that as $\theta \to 0$, $ar_0 \to 0$ and as $\theta$ becomes large
$ar_0$ becomes large. So, $ar_0$ is a measure of non-commutativity. Now
substituting \eqn{parameter1}, \eqn{parameter2} and \eqn{ar} in
\eqn{dragforcesmall}, we obtain for small non-commutativity (i.e. 
$ar_0 \ll 1$), the expression of the drag force as,
\be\label{dragforcesmall1}
F = \frac{1}{2\pi} \frac{v}{\sqrt{1-v^2}}\frac{(4\pi T)^{\frac{7-p}{5-p}}
d_p^{\frac{1}{5-p}}\hat{\lambda}^{\frac{1}{5-p}}}{(7-p)^{\frac{7-p}{5-p}}}
\left[1-\frac{2-v^2}{2(1-v^2)^2}\frac{\theta^2 (4\pi T)^{\frac{2(7-p)}{5-p}}
d_p^{\frac{2}{5-p}} \hat{\lambda}^{\frac{2}{5-p}}}{(7-p)^{\frac{2(7-p)}{5-p}}}
+ \cdots\right]
\ee        
where `dots' represent terms of the order 
$(T^{(7-p)/(5-p)} \hat{\lambda}^{1/(5-p)} \theta)^4$ and
higher. It is clear that when the non-commutativity parameter $\theta$
is set to zero we recover the drag force expression for the 
$(p+1)$-dimensional ordinary YM theory \cite{Liu:2006he}. 

Next substituting \eqn{parameter1}, \eqn{parameter2} and \eqn{ar} in 
\eqn{dragforcebig}, we obtain the expression of the drag force for large 
non-commutativity (i.e. $ar_0 \gg 1$) as,
\be\label{dragforcebig1}
F = \frac{v}{2\pi} \frac{(7-p)^{\frac{7-p}{5-p}}}{(4\pi T)^{\frac{7-p}{5-p}}
d_p^{\frac{1}{5-p}}\hat{\lambda}^{\frac{1}{5-p}}\theta^2}
\left[1-\frac{2+v^2}{2}\frac{(7-p)^{\frac{2(7-p)}{5-p}}}
{\theta^2 (4\pi T)^{\frac{2(7-p)}{5-p}}
d_p^{\frac{2}{5-p}} \hat{\lambda}^{\frac{2}{5-p}}}
+ \cdots\right]
\ee
here `dots' represent terms of the order 
$(T^{(7-p)/(5-p)} \hat{\lambda}^{1/(5-p)} \theta)^{-4}$ and higher. We point 
out that
the dependence of temperature and 't Hooft coupling in the expressions of drag
force \eqn{dragforcebig1} for large non-commutativity got inverted from that 
of small non-commutativity \eqn{dragforcesmall1}. Also for both small and
large non-commutativity the drag force gets reduced and so, the viscosity
of the plasma gets reduced by the effect of non-commutativity. This was also 
observed for the case of (3+1)-dimensional NCYM theory \cite{Matsuo:2006ws}. 
However, we find that
the expression for the drag force in general depends on $p$, and so there 
is no universal structure in the drag force expression for the general 
$(p+1)$-dimensional theory. However, we will mention, as was also noted in 
\cite{Liu:2006he} for the jet quenching parameter, that we can write the 
drag force in
terms of some quantity to uncover a kind of
`universal structure' in the drag force expression for all $(p+1)$-dimensional
gauge theory. For this, let us note that the entropy density of the
non-extremal (D$(p-2)$, D$p$) in the decoupling limit can be obtained
from \eqn{newmetric} as,
\be\label{entropy}
s = \frac{4\pi\Omega_{8-p} r_0^{8-p} \sinh\varphi}{16 \pi G \hat g^2}
\ee
where $G$ is the Newton's constant in ten dimensions and $16\pi G = (2\pi)^7
\alpha'^4$. The entropy density of
non-extremal (D$(p-2)$, D$p$) is known to have the same form as ordinary 
non-extremal D$p$-branes, but
we have put $\hat g$ as the string coupling in the decoupling limit as opposed
to the original string coupling $g$. They are related as, $g =
(\a'/\theta)\hat g$, with $\hat g$ kept fixed \cite{Maldacena:1999mh}. 
The reason for the difference
is that the entropy density $s$ in the lhs also contains a volume which is 
scaled (due to non-commutativity) by the same factor. Now substituting 
$r_0$, $\sinh\varphi$ from
\eqn{parameter1} and \eqn{parameter2} and using $\hat{g}_{YM}^2 = (2\pi)^{p-2}
\hat g \alpha'^{(p-3)/2}$, we get from \eqn{entropy},
\be\label{entropy1}
s = N^2 \hat{\lambda}^{\frac{p-3}{5-p}} T^{\frac{9-p}{5-p}} b_p
\ee
where 
\be\label{bp}
b_p = \left[\frac{2^{16-3p} \pi^{\frac{13-3p}{2}}
    \Gamma\left(\frac{7-p}{2}\right)}{(7-p)^{7-p}}\right]^{\frac{2}{5-p}}
\ee      
Introducing a dimensionless 't Hooft coupling by $\hat\lambda_{\rm eff}(T) =
T^{p-3} \hat\lambda$, we can rewrite \eqn{entropy1} as,
\be\label{entropy2}
s = N^2 b_p \hat\lambda_{\rm eff}^{\frac{p-3}{5-p}}(T) T^p \equiv 
N^2 c(\hat\lambda, T) T^p
\ee
So, it is clear that $c \equiv b_p \hat\lambda_{\rm eff}^{\frac{p-3}{5-p}}$ 
characterizes the number of degrees of freedom of $(p+1)$-dimensional 
NCYM theory at temperature $T$ \cite{Liu:2006he}. We can rewrite the drag 
force expression
\eqn{dragforcesmall1} (for small non-commutativity or $\theta \ll 1$) in terms 
of $c$ and $\hat\lambda_{\rm eff}$ as
\be\label{univdragforce1}
F = \frac{v}{\sqrt{1-v^2}} c^{\frac{1}{2}}(\hat\lambda, T) 
\sqrt{\hat\lambda_{\rm eff}(T)} T^2 \left[1 - \frac{2-v^2}{2(1-v^2)^2} 4\pi^2
  \theta^2
c(\hat\lambda,T) \hat\lambda_{\rm eff}(T) T^4 + \cdots\right]
\ee
Similarly we can rewrite the drag force expression
\eqn{dragforcebig1} (for large non-commutativity or $\theta \gg 1$) as
\be\label{univdragforce2}
F = \frac{v}{4\pi^2\theta^2} c^{-\frac{1}{2}}(\hat\lambda, T) 
\frac{1}{\sqrt{\hat\lambda_{\rm eff}(T)} T^2} \left[1 - \frac{2+v^2}{2} 
\frac{1}{4\pi^2\theta^2}
c(\hat\lambda,T)^{-1} \frac{1}{\hat\lambda_{\rm eff}(T) T^4} + \cdots\right]
\ee
When $p=3$, the constant $c$ reduces to $\pi^2/2$ and in that case our
results match with those given in ref.\cite{Matsuo:2006ws} except for a 
factor of 2 
(our definition of $r_0^{7-p}\sinh^2\varphi$ differs from that
in ref.\cite{Matsuo:2006ws} by a factor of 2). It is thus clear that 
writing in terms of
an effective 't Hoooft coupling $\hat\lambda_{\rm eff}$ and 
$c(\hat\lambda, T)$, a quantity characterizing the number of degrees of
freedom at $T$, the drag force expression can be expressed into a `universal'
form and $c$ encodes the `non-universal' structure.

To summarize, in this note we use holography and fundamental string 
as a probe in the background of non-extremal (D$(p-2)$, D$p$) brane bound 
state system in a particular decoupling limit to calculate the drag force on 
an external quark moving in a hot NCYM plasma in various dimensions. We find
that when the quark moves in one of the non-commutative directions the drag
force gets reduced and thus non-commutativity makes the plasma less viscous.
We have calculated the drag force for both small non-commutativity and large
non-commutativity. For small non-commutativity if we set the non-commutative
parameter to zero we recover the results for the commutative theory as
expected. For large non-commutativity, the dependence of the temperature and
the 't Hooft coupling on the drag force gets inverted from that of the small
non-commutativity. These results were known for (3+1)-dimensional NCYM theory,
and we calculated the same for the general $(p+1)$-dimensional NCYM theory
and we do not find any universal structure in the expression of the drag
force in general. However, a `universal structure' emerges if we write the
drag force in terms of a quantity ($c$) which characterizes the number of 
degrees of freedom in the gauge theory and appear in the entropy density.
It would be interesting to understand the true physical meaning of this 
quantity in the non-commutative gauge theories.


\end{document}